\documentclass{article}
\usepackage[english]{babel}
\usepackage[letterpaper,top=2cm,bottom=2cm,left=3cm,right=3cm,marginparwidth=1.75cm]{geometry}
\usepackage{amsmath}
\usepackage{graphicx}
\usepackage[colorlinks=true, allcolors=blue]{hyperref}
\usepackage{subcaption}
\usepackage{float}
\usepackage{indentfirst}
\usepackage[numbers,sort&compress]{natbib}

\title{Adaptive network dynamics and behavioral contagion in multi-state drug use propagation}

\author{
Hsuan-Wei Lee$^{1*}$, Yi-Hsuan Huang$^2$, Nishant Malik$^{3}$ \\
\\
$^1$College of Health, Lehigh University, USA \\
$^2$Department of Public Finance, National Chengchi University, Taiwan \\
$^3$School of Mathematics and Statistics, Rochester Institute of Technology, USA \\
$^*$Corresponding author: hsl324@lehigh.edu
}

\date{}

\begin{document}
\maketitle

\begin{abstract}
Addictive behavior spreads through social networks via feedback among choice, peer pressure, and shifting ties, a process that eludes standard epidemic models. We present a comprehensive multi-state network model that integrates utility-based behavioral transitions with adaptive network rewiring, capturing the co-evolutionary dynamics between drug use patterns and social structure. Our framework distinguishes four distinct individual states by combining drug use behavior with addiction status, while allowing individuals to strategically disconnect from drug-using neighbors and form new connections with non-users. Monte Carlo simulations show that rewiring reshapes contagion, pulling high-degree nodes into drug-free clusters and stranding users on sparse fringes. Systematic exploration of the four-dimensional parameter space reveals sharp phase transitions reminiscent of critical phenomena in statistical physics, where small changes in recovery rates or addiction conversion rates trigger dramatic shifts in population-level outcomes. Most significantly, the rewiring probability emerges as the dominant control parameter, establishing adaptive network management as more influential than biological susceptibility factors in determining addiction prevalence. Our findings challenge traditional intervention paradigms by revealing that empowering individuals to curate their social environments may be more effective than targeting individual behavioral change alone.
\end{abstract}


\section{\textbf{Introduction}}

Addictive behaviours circulate through social networks in ways that challenge classical contagion theory \cite{friedman2006addictive, rosenquist2010spread, thombs2019introduction}. Their spread depends on deliberate cognition rather than on the passive exposure that drives infectious disease \cite{khantzian2003understanding, bechara2019neurobehavioral, hunt2024evolutionary}. Pathogens travel passively and exploit molecular interactions, but behaviours move only when people choose. In stark contrast, behavioral epidemics emerge from strategic decision-making processes \cite{masaeli2021prevalence, alimoradi2022estimation}. People weigh short-term reward against long-term harm while their social ties shift. This distinction exposes critical limitations in traditional epidemiological frameworks. These models assume fixed network topologies and probabilistic transmission mechanisms. Standard epidemic models fail to capture this volitional adoption-and-cessation cycle.

Conventional compartmental models overlook three core traits that set behavioural contagion apart from infection \cite{allen1994some, tuckwell2007some, battista2019modeling, tolles2020modeling, mayengo2020mathematical, van2022review}. First, drug use stems from explicit utility calculations. These calculations incorporate multifaceted cost-benefit trade-offs that extend far beyond simple exposure probabilities. Second, addiction progression unfolds through discrete neurobiological phases. The trajectory encompasses experimental use, physiological dependence, and potential recovery phases, which binary state classifications cannot capture. Third, social networks exhibit extraordinary plasticity. Individuals actively curate relationships by severing threatening connections while cultivating and reinforcing bonds based on behavioral compatibility. These fundamental differences demand mathematical frameworks with unprecedented capabilities. Such models must simultaneously capture individual agency in decision-making processes, the multi-stage nature of addiction progression, and the dynamic evolution of social network topology through strategic relationship management. Adaptive-network studies already show that co-evolving states and links steer collective outcomes \cite{huang2014interplay, karakose2023comprehensive, zhao2024investigating, nj23101968}. Studies reveal how strategic rewiring mechanisms can fundamentally alter behavioral transmission patterns and emergent network structures in evolving social systems \cite{marceau2010adaptive, malik2016transitivity, lee2018evolutionary, nmrole2013}.

Empirical investigations reveal that these adaptive network processes generate counterintuitive population-level phenomena. Protective clustering mechanisms spontaneously create resilient communities for healthy individuals \cite{kirmayer2009community, johnson2015infectious, bardosh2017addressing}. Simultaneously, these exact mechanisms isolate vulnerable populations in sparse subnetworks. These isolated networks are characterized by diminished social capital and severely limited recovery resources. When neurobiological addiction processes outpace social adaptation capabilities, however, behavioral epidemics can breach even well-fortified protective clusters \cite{choi2019neurobiological, valentino2020neurobiology, volkow2023substance}. This breach precipitates system-wide entrenchment that transforms localized outbreaks into endemic conditions through cascading failure mechanisms. The dynamic tension between competing forces creates complex phase spaces. In these systems, infinitesimal parameter variations can trigger catastrophic shifts in population outcomes. These phenomena exhibit characteristics reminiscent of critical phenomena in statistical physics \cite{di2012statistical, d2015statistical, crokidakis2021modeling, bergstrom2024human}. Yet existing theoretical frameworks lack the mathematical sophistication necessary to predict or explain these sharp transitions. These transitions determine the fate of entire populations. The emergence of social clustering in epidemic spread on coevolving networks creates protective barriers that fundamentally alter disease transmission dynamics \cite{grabowski2012relationship, lee2019social, nunner2022health}. The incorporation of spatial heterogeneity and interactive diversity can further modify adaptive behaviors and collective health outcomes \cite{lloyd1996spatial, real2007spatial, xiong2024interacting, si2025evolution}.

We address this theoretical gap by constructing a comprehensive mathematical framework. Our approach seamlessly integrates utility-based decision-making with adaptive network dynamics and multi-state neurobiological progression. This creates the first model capable of capturing the full complexity of behavioral epidemics in evolving social systems. We transcend binary behavioral classifications by implementing a sophisticated four-state taxonomy. This framework captures the critical cross-classification between drug use behavior and addiction status. The approach distinguishes between using and non-using behavior and between addicted and potential-addicted status. This enables a more realistic representation of addiction trajectories while preserving analytical tractability essential for large-scale computational exploration. The decision-making architecture employs meticulously calibrated utility functions. These functions explicitly account for peer influence gradients \cite{henneberger2021peer}, withdrawal symptom penalties \cite{duncan2019fast}, baseline behavioral preferences \cite{field2020recovery}, and social stigma effects \cite{matthews2017stigma}. Modified Fermi dynamics govern state transitions \cite{traulsen2009stochastic, altrock2009deterministic}. These dynamics transform utility differentials into smooth probability distributions. The distributions reflect empirically documented decision-making patterns under neurobiological constraints. This utility-based approach builds upon recent advances in granular learning algorithms. These algorithms demonstrate how individual adaptation strategies can enhance collective welfare through sophisticated decision-making mechanisms \cite{robert2016processes, lee2025granular}.

The revolutionary component of our framework lies in its incorporation of adaptive rewiring mechanisms. These mechanisms allow network topology to co-evolve with behavioral dynamics. This approach reflects the empirically established tendency for individuals to probabilistically sever connections with behaviorally incompatible neighbors. Simultaneously, individuals form new relationships within preferred behavioral communities through homophilic selection processes. The co-evolutionary coupling between individual choices and network structure generates emergent phenomena. These phenomena fundamentally alter transmission dynamics compared to static network models. The process produces degree-stratified segregation where social connectivity becomes strongly correlated with behavioral status. This correlation emerges through purely local rewiring decisions. These strategies create self-organizing structures that promote beneficial collective outcomes \cite{tortia2025stakeholders, lee2025enhancing}. This provides empirical support for the transformative potential of adaptive network management in complex social systems. Our systematic exploration encompasses a four-dimensional parameter space. The space is defined by peer influence strength ($\delta_s$), addiction stickiness ($\delta_w$), recovery rates ($\alpha$), addiction conversion rates ($\beta$), and rewiring probabilities ($\gamma$). We conduct comprehensive Monte Carlo simulations spanning $11^4 = 14,641$ parameter combinations. These simulations operate across networks containing up to $10,000$ nodes to ensure statistical robustness. The approach identifies universal principles governing behavioral epidemic dynamics. Through this exhaustive computational investigation, we map the complete phase diagram of possible behavioral outcomes. We pinpoint critical transition boundaries where targeted interventions can achieve maximum therapeutic leverage. These interventions operate through nonlinear amplification effects. Our analysis reveals that the rewiring parameter exerts dominant control over long-term addiction prevalence. This parameter completely overshadows biological susceptibility factors in determining population health trajectories.

Our theoretical framework provides a versatile foundation for understanding diverse behavioral phenomena. These phenomena occur where strategic choice, social influence, and network adaptation converge to produce collective outcomes. Applications span domains from political polarization and misinformation propagation to health behavior diffusion and technological innovation adoption \cite{baldassarri2007dynamics, liu2018investigating, ahmad2023households, lu2024agents}. The model's demonstrated capacity to predict sharp phase transitions and identify leverage points for intervention makes it particularly valuable. This capability enables the design of evidence-based public health strategies that harness rather than oppose natural social dynamics. The mathematical rigor enables the precise quantification of intervention effectiveness across diverse population contexts. Beyond immediate applications to addiction research, our work advances fundamental understanding of complex adaptive systems. We demonstrate how local decision-making rules generate emergent global patterns that qualitatively transform system behavior. The framework offers insights into the mathematical principles that govern self-organization in biological and social networks. Individual adaptive capacity can amplify into systemic advantage or disadvantage through network effects. These effects operate independently of explicit discrimination or resource allocation mechanisms.

The mathematical techniques we employ bridge rational choice theory with stochastic network evolution. This establishes a methodological template for investigating multi-scale phenomena. These phenomena occur where individual agency and collective organization interact to produce complex population-level dynamics across diverse scientific domains. The remainder of this paper systematically presents our contributions across three main sections. Section 2 details the mathematical formulation of our multi-state adaptive network model. This encompasses utility-based decision dynamics and neurobiological state transitions. Section 3 provides comprehensive simulation results. These results reveal temporal evolution patterns, emergent network reorganization, and systematic parameter sensitivity analysis across the complete four-dimensional parameter space. Section 4 offers a detailed discussion of implications for both theoretical understanding and practical intervention design in complex adaptive systems. We identify promising directions for future research. These directions could extend our framework to incorporate heterogeneous individual characteristics, empirical network topologies, and co-diffusing behavioral phenomena. Such extensions would create even more sophisticated representations of real-world social systems.

\section{\textbf{Methods}}

\subsection{\textit{Mathematical framework for behavioral contagion in adaptive networks}}

We build a deterministic framework that links individual decisions with a network that rewires over time. The model couples neurobiological addiction progression with social rewiring, so behavior and topology evolve together. The foundation of our model is a time-evolving random network $G(t) = (V, E(t))$ consisting of $N$ nodes representing individuals, where the edge set $E(t)$ undergoes adaptive rewiring based on behavioral compatibility. Each node $i \in V$ maintains a time-varying degree $k_i(t)$, with the network characterized by its degree distribution $\{p_k(t)\}$, where $p_k(t)$ represents the probability that a randomly selected node has exactly $k$ connections at time $t$.

The initial network topology follows a Poisson degree distribution to ensure analytical tractability while capturing the essential features of random social networks:
\begin{equation}
p_k(0) = \frac{\langle k \rangle^k e^{-\langle k \rangle}}{k!}
\end{equation}
where $\langle k \rangle$ denotes the mean degree. Self-loops and parallel edges are disallowed, which keeps $\sum_i k_i(t)$ even and the graph valid. Our model transcends binary state representations by incorporating a four-state classification that captures the complex neurobiological progression of addiction. The four states mirror clinical evidence that addiction has stages, which binary $S$-$I$ models cannot capture \cite{lee2019social, bechara2019neurobehavioral}. Each individual $i$ is characterized by two independent attributes that define their behavioral and physiological status: (1) Behavioral State $b_i(t) \in \{0, 1\}$ indicating non-drug-using ($b_i = 0$) or drug-using ($b_i = 1$) behavior, and (2) Addiction Status $a_i(t) \in \{0, 1\}$ representing potential-addicted ($a_i = 0$) or addicted ($a_i = 1$) condition. This classification yields four distinct biological states: non-drug-using potential-addicted (NP: $b_i = 0, a_i = 0$), drug-using potential-addicted (DP: $b_i = 1, a_i = 0$), non-drug-using addicted (NA: $b_i = 0, a_i = 1$), and drug-using addicted (DA: $b_i = 1, a_i = 1$). The initial condition reflects empirical observations of addiction onset from longitudinal studies, with 1\% of the population randomly assigned to drug-using states and all individuals initialized as potentially addicted, representing the early stages of an emerging behavioral epidemic. This initialization protocol aligns with epidemiological evidence showing that behavioral epidemics typically begin with small fractions of early adopters before spreading through social networks \cite{rosenquist2010spread}.

\subsection{\textit{Utility-based decision dynamics and neurobiological transitions}}

Each choice follows a utility rule that combines peer pressure, withdrawal cost, and baseline preference. This utility-based approach is grounded in rational choice theory, which has been successfully applied to model decision-making in addiction contexts where individuals weigh costs and benefits despite neurobiological constraints \cite{bechara2019neurobehavioral}. We construct state-dependent utility functions that capture the distinct neurobiological contexts faced by addicted versus potential-addicted individuals. For an addicted individual $i$ with addiction status $a_i = 1$, the utilities for drug use ($D$) and non-use ($N$) incorporate withdrawal effects and social influences:
\begin{align}
U^{A}_{Di} &= \eta + \theta \cdot \delta_s \cdot \rho_i^{\theta} - \omega_i \label{eq:utility_AD}\\
U^{A}_{Ni} &= \eta + \theta \cdot \delta_s \cdot (1 - \rho_i^{\theta}) - \delta_w \label{eq:utility_AN}
\end{align}
For a potential-addicted individual $j$ with addiction status $a_j = 0$:
\begin{align}
U^{P}_{Dj} &= \eta + \theta \cdot \delta_s \cdot \rho_j^{\theta} - \omega_j \label{eq:utility_PD}\\
U^{P}_{Nj} &= \eta + \theta \cdot \delta_s \cdot (1 - \rho_j^{\theta}) \label{eq:utility_PN}
\end{align}
The parameter $\rho_i = \frac{1}{k_i} \sum_{j \in \mathcal{N}_i} b_j$ represents the proportion of drug users among individual $i$'s neighbors $\mathcal{N}_i$, capturing the local social environment. This formulation reflects empirical evidence that peer influence in substance use operates primarily through local social networks, with the strength of influence proportional to the density of users in an individual's immediate social circle \cite{rosenquist2010spread}. We fix the baseline $\eta = 1$ and set $\theta = 0.7$ to temper the influence of peers. The deterrence penalty $\omega_i$ captures social stigma effects when individuals have no drug-using neighbors:
\begin{equation}
\omega_i = \begin{cases}
0.1 \cdot k_i & \text{if } \rho_i = 0 \\
0 & \text{otherwise}
\end{cases}
\end{equation}
The withdrawal penalty $\delta_w$ represents the neurobiological costs of abstinence for addicted individuals, reflecting documented effects of neuroadaptation, while $\delta_s$ quantifies the strength of peer influence effects on behavioral decisions. These neurobiological parameters are calibrated to reflect clinical understanding of withdrawal symptoms and peer influence mechanisms documented in addiction neuroscience research \cite{volkow2023substance}.

Modified Fermi functions transform utility gaps into transition odds, capturing the asymmetry between quitting and relapse. The core Fermi function $\frac{1}{1 + \exp(\Delta U)}$ provides the foundation for rational decision-making based on utility differences, but addiction involves systematic deviations from pure rationality due to neurobiological changes that require explicit modeling. The Fermi function approach has been widely used in evolutionary game theory and opinion dynamics to model probabilistic decision-making under uncertainty, providing a mathematically tractable framework for capturing bounded rationality in social systems \cite{szabo2005phase, perc2010coevolutionary, lee2018evolutionary}. The transition probabilities are constructed as:
\begin{align}
\phi(D \to N) &= \frac{1}{1 + \exp(U_D - U_N)} \cdot (1 - \tau \cdot \rho_i^{\theta}) \label{eq:prob_DN}\\
\phi(N \to D) &= \frac{1}{1 + \exp(U_N - U_D)} \cdot (\lambda + (1 - \lambda) \cdot \rho_i^{\theta}) \label{eq:prob_ND}
\end{align}
The multiplicative modifications capture two fundamental asymmetries in addiction neurobiology. For cessation attempts (Equation~\ref{eq:prob_DN}), the factor $(1 - \tau \cdot \rho_i^{\theta})$ represents social pressure suppression, where higher proportions of drug-using neighbors systematically reduce the probability of quitting through mechanisms including peer pressure, cue exposure, and reduced access to recovery role models. For initiation attempts (Equation~\ref{eq:prob_ND}), the factor $(\lambda + (1 - \lambda) \cdot \rho_i^{\theta})$ ensures a baseline probability $\lambda = 0.3$ of behavioral change independent of social context, reflecting autonomous decision-making capacity. At the same time, the peer influence component scales the remaining probability based on social exposure. The parameter $\tau = 0.5$ modulates the strength of social pressure suppression, calibrated to clinical observations of relapse triggers. These asymmetric transition probabilities reflect established findings in addiction psychology that cessation attempts face more substantial social barriers than initiation, particularly in environments with high user density \cite{hunt2024evolutionary}.

Beyond behavioral choices, individuals undergo physiological transitions that reflect the biological progression of addiction according to established neurobiological principles. These transitions operate independently of the utility-based behavioral dynamics, capturing the medical reality that addiction status can change even when behavior remains constant. The evolution of addiction status follows Markovian dynamics with state-dependent transition rates: $P(a_i(t+1) = 1 | a_i(t) = 0, b_i(t) = 1) = \beta$ representing addiction conversion through repeated drug exposure, and $P(a_i(t+1) = 0 | a_i(t) = 1, b_i(t) = 0) = \alpha$ representing neurobiological recovery during abstinence. This formulation reflects clinical observations that addiction develops through neuroadaptation from repeated drug exposure and can be reversed through sustained abstinence, though underlying vulnerability may persist. This Markovian approach to addiction progression aligns with current understanding of neurobiological changes in addiction, where repeated exposure leads to measurable alterations in brain structure and function that can be partially reversed through sustained abstinence \cite{valentino2020neurobiology}.

\subsection{\textit{Adaptive network rewiring and computational implementation}}

The distinguishing feature of our model is the incorporation of adaptive rewiring, which allows the network topology to co-evolve with behavioral dynamics, reflecting empirical observations that individuals actively manage social connections based on behavioral compatibility. Rewiring occurs probabilistically when behavioral discordance exists between connected individuals. For any edge $(i,j) \in E(t)$ where $b_i(t) \neq b_j(t)$, the link is severed with probability $\gamma$ per time step, with rewiring initiated by the non-drug-using individual based on empirical evidence that healthy individuals are more motivated to avoid harmful social influences. Upon severing the link, the non-drug-using node immediately establishes a new connection to a randomly selected non-drug-using individual from the population, subject to constraints that prevent self-connections and duplicate links. This preferential attachment to behaviorally similar individuals drives the emergence of homophilic clustering while maintaining constant network density. The rewiring mechanism is formally described as:
\begin{equation}
P(\text{rewire}(i,j)) = \begin{cases}
\gamma & \text{if } b_i \neq b_j \text{ and } b_i = 0 \\
0 & \text{otherwise}
\end{cases}
\end{equation}
This asymmetric rewiring rule captures the empirically supported tendency for individuals to disengage from relationships that conflict with their behavioral identity while seeking connections that reinforce positive behaviors.
\begin{figure}[h]
\centering
\includegraphics[width=0.35\linewidth]{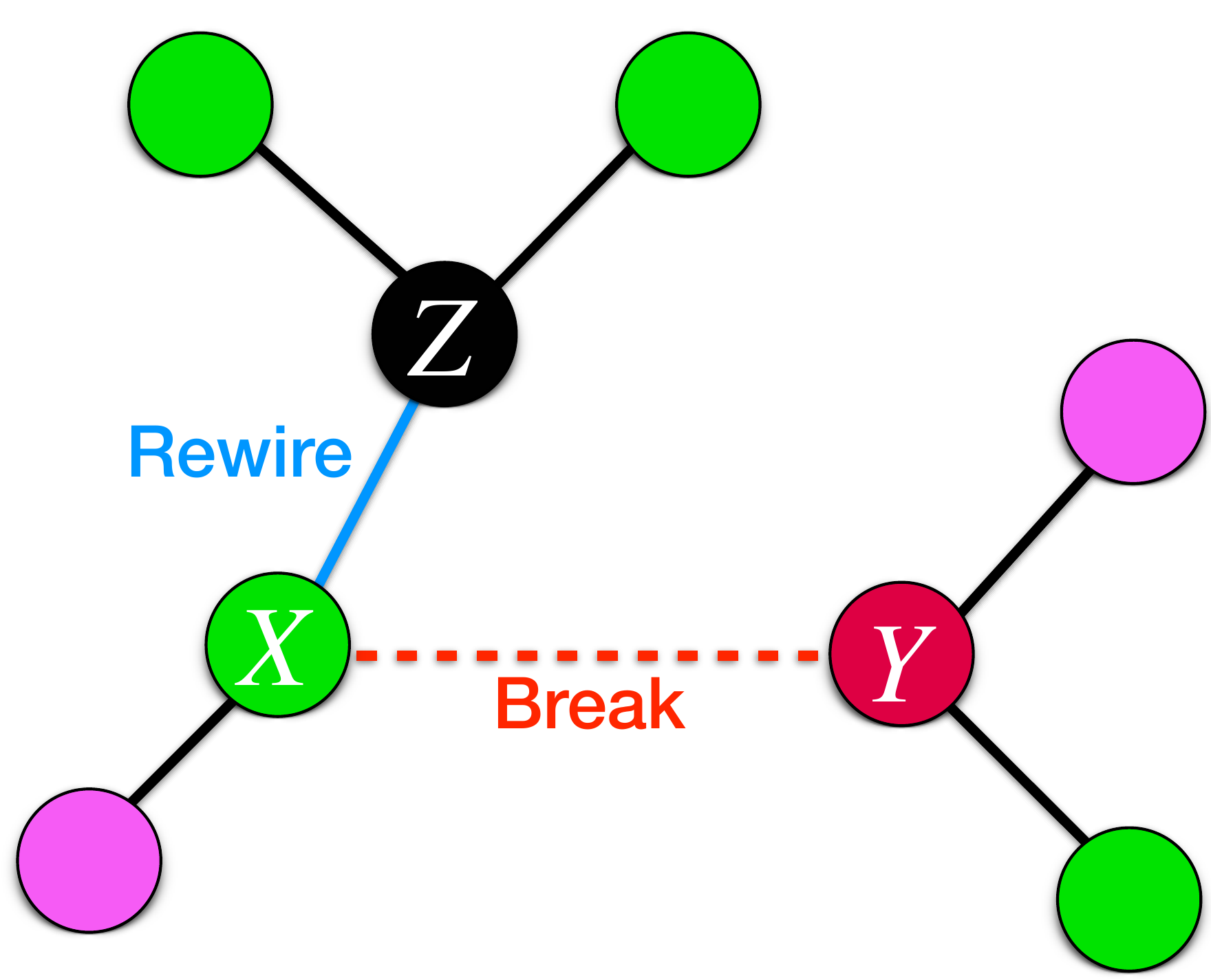}
\caption{\label{fig:Rewiring}Illustration of adaptive network rewiring. Different colors represent the four biological states: NP (green), DP (pink), NA (black), and DA (red). A discordant edge between node $X$ (NP state) and node $Y$ (DA state) undergoes rewiring with probability $\gamma$, where the non-drug-using node $X$ severs the connection and forms a new link with a randomly selected non-drug-using node $Z$.}
\end{figure}

Our computational framework employs Monte Carlo simulations with synchronous updating to ensure temporal consistency across all biological and social processes. Each simulation time step consists of three sequential phases reflecting natural temporal hierarchies: (1) utility-based behavioral transitions according to neurobiological decision-making principles, (2) physiological state evolution based on addiction conversion and recovery processes, and (3) adaptive network rewiring driven by social compatibility preferences. We conduct a systematic exploration of the four-dimensional parameter space $(\delta_s, \delta_w, \alpha, \beta) \in [0,1]^4$ using comprehensive grid sampling with $11^4 = 14,641$ parameter combinations, treating rewiring probability $\gamma$ as an independent control parameter. Each parameter configuration involves multiple independent simulations with different random seeds to quantify statistical variability and ensure robust conclusions.

The system is initialized with 1\% drug users randomly distributed across the network, with all individuals in potential-addicted states, reflecting realistic epidemic initiation conditions. Each run lasts $10^{4}$ Monte Carlo steps, long enough for the key observables to stabilize. Network sizes range from $N = 2,000$ to $N = 10,000$ nodes to verify that observed phenomena represent genuine emergent properties rather than finite-size artifacts, with mean degree held constant at $\langle k \rangle = 2$ to maintain consistent local interaction intensity. Our implementation utilizes efficient algorithms for dynamic network manipulation and vectorized utility calculations, with standardized random number generation protocols ensuring reproducibility. Results are aggregated across multiple independent runs to provide robust estimates of steady-state measures and confidence intervals, enabling comprehensive characterization of the behavioral phase diagram and identification of critical transition boundaries where small parameter changes produce dramatic shifts in population-level outcomes.

\section{\textbf{Results}}

We report simulations that track how choice and network structure co-evolve during behavioral contagion. Our systematic exploration encompasses temporal dynamics, emergent network reorganization, and parameter sensitivity across the four-dimensional space of peer influence ($\delta_s$), addiction stickiness ($\delta_w$), recovery rates ($\alpha$), addiction conversion rates ($\beta$), and rewiring probabilities ($\gamma$). All simulations begin with 1\% drug users randomly distributed across initially random networks, with all individuals in potential-addicted states, reflecting realistic epidemic initiation conditions observed in longitudinal studies of substance use onset.

\subsection{\textit{Emergent temporal dynamics and spontaneous network self-organization}}

\begin{figure}[h]
\centering
\includegraphics[width=0.7\linewidth]{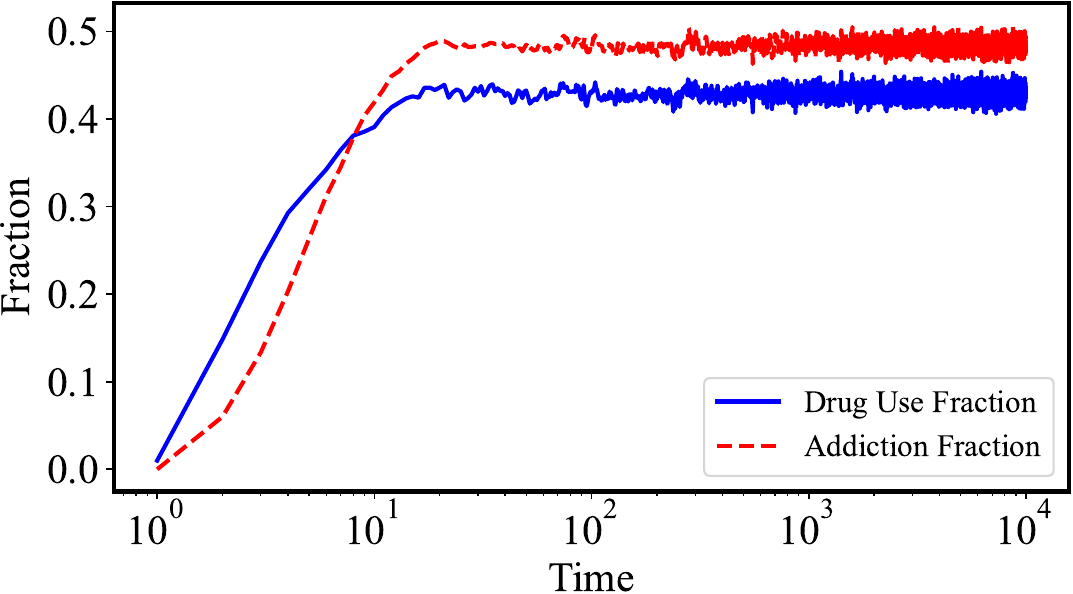}
\caption{\label{fig:TimeEvolution}Temporal evolution of drug use and addiction fractions in an adaptive network system. Drug use (solid lines) and addiction (dashed lines) fractions evolve over time $t$ on a network with mean degree $\langle k \rangle = 2$ and population size $N = 10,000$. The simulation demonstrates the characteristic two-phase dynamics with parameters $\delta_s = 0.4$, $\delta_w = 0.9$, $\alpha = 0.4$, $\beta = 0.4$, and $\gamma = 0.6$, showing initial epidemic spread followed by network-mediated stabilization.}
\end{figure}

The time series exhibits two distinct phases that are not evident in static networks. Figure~\ref{fig:TimeEvolution} demonstrates this dynamic progression for representative parameter values, revealing the intricate interplay between behavioral spreading and network adaptation over time. The initial phase (approximately $t < 10^2$) is characterized by rapid behavioral contagion where drug use spreads through the network via peer influence mechanisms, closely resembling traditional epidemic dynamics in static networks. During this period, both drug use and addiction fractions increase exponentially as behavioral transmission dominates over network adaptation effects. The addiction fraction systematically lags behind drug use prevalence, reflecting the finite time required for neurobiological dependence to develop following initial drug exposure according to the addiction conversion rate $\beta$, consistent with clinical observations of addiction progression.

The transition to the second phase occurs when network rewiring effects accumulate sufficiently to alter the fundamental transmission dynamics, demonstrating how biological systems can spontaneously develop protective mechanisms through social adaptation. As non-drug users increasingly sever connections with drug-using neighbors and form new links within the non-using population, the effective connectivity between behavioral groups diminishes, creating natural barriers to further spread. Rewiring pushes the system to a steady state where spread and rewiring balance. The convergence to steady state at approximately $t = 10^2$ represents a fundamental shift from epidemic spreading to endemic equilibrium, where the population achieves a dynamically stable configuration of behaviors and network structure that demonstrates the efficiency of adaptive network mechanisms in establishing protective population-level organization against behavioral contagion.

\begin{figure}[h]
\centering
\includegraphics[width=0.45\linewidth]{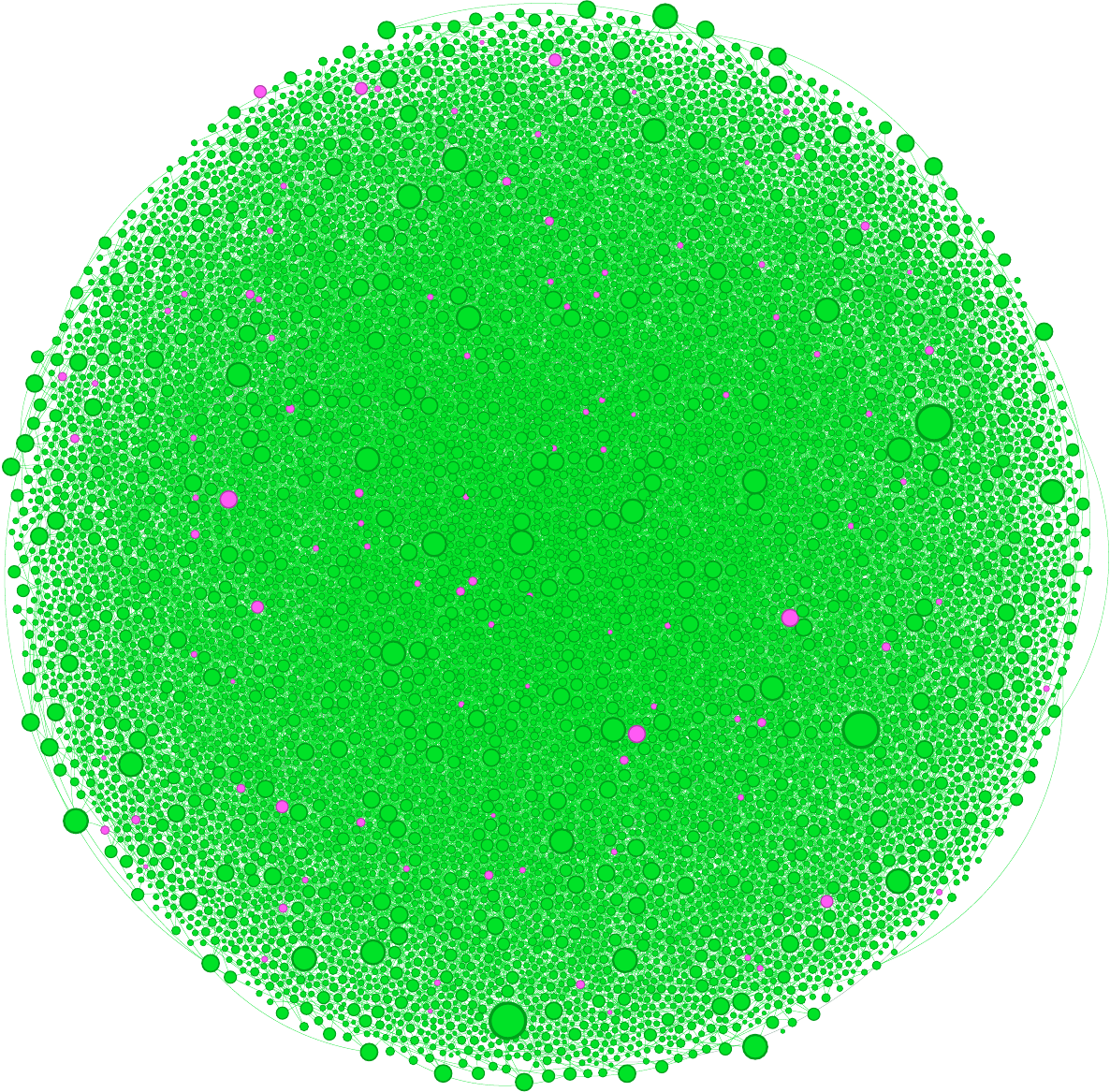}
\includegraphics[width=0.45\linewidth]{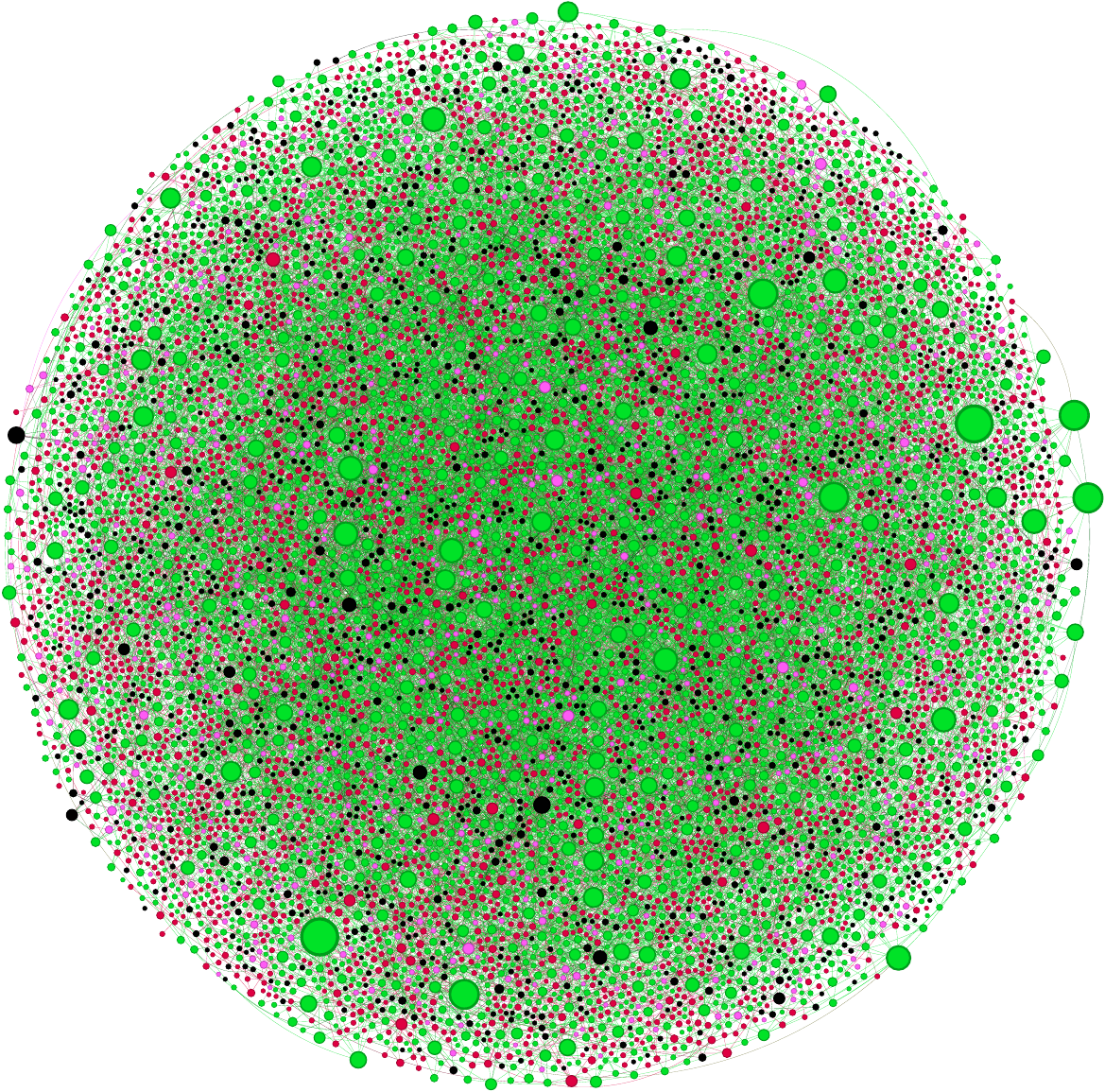}
\caption{\label{fig:NetworkStates}Network visualization demonstrating spontaneous degree-stratified segregation. The four biological states are color-coded: non-drug-using potential-addicted (NP, green), drug-using potential-addicted (DP, pink), non-drug-using addicted (NA, black), and drug-using addicted (DA, red). Node size represents degree connectivity. The left panel shows the initial random configuration; the right panel shows the steady-state organization after $10^4$ time steps. Network parameters: $N=10,000$, $\delta_s=0.4$, $\delta_w=0.9$, $\alpha=0.4$, $\beta=0.4$, $\gamma=0.6$. Visualizations created using ForceAtlas 2 layout in Gephi.}
\end{figure}

A key outcome is degree-stratified clustering that links connectivity to risk. Figure~\ref{fig:NetworkStates} provides a vivid illustration of this transformation, contrasting the initial random configuration with the highly organized steady-state structure that emerges purely through local rewiring decisions. Initially, the network exhibits homogeneous mixing with $99\%$ non-drug-using potential-addicted (NP) individuals and $1\%$ drug-using potential-addicted (DP) individuals distributed randomly throughout the structure, with node sizes representing degree centrality showing no correlation with behavioral status. This baseline confirms that any subsequent stratification represents genuine emergent behavior rather than initialization artifacts.

The steady-state configuration presents a dramatically different landscape where degree and behavioral status exhibit strong correlation, creating a form of emergent social stratification based on health status. Large, highly connected nodes predominantly occupy non-drug-using states (both NP and NA categories), while small, poorly connected nodes cluster in drug-using categories (both DP and DA states). Many drug-using nodes become completely isolated, trapped in behavioral states with minimal social connectivity that severely limits their access to social support and recovery resources. This degree stratification emerges without any explicit degree-based preferences in the rewiring mechanism, arising solely from the differential effects of behavioral homophily on network connectivity, and represents a powerful example of self-organization in biological systems.

\begin{figure}[H]
\centering
\includegraphics[width=1\linewidth]{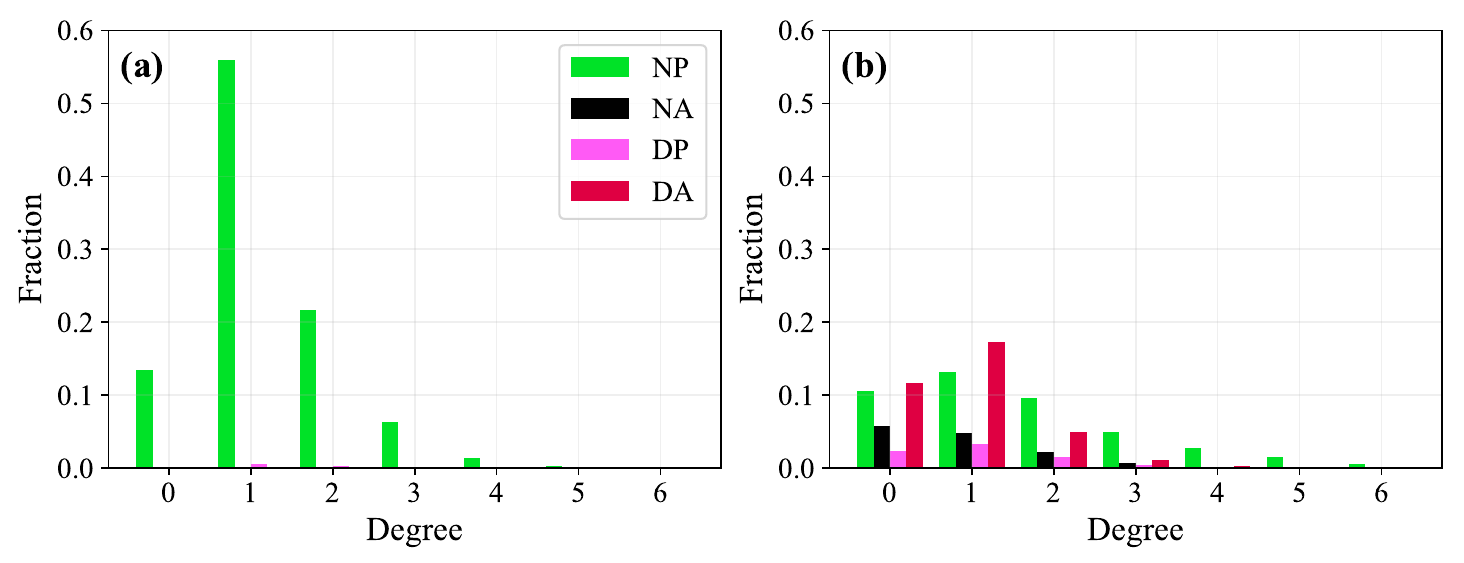}
\caption{\label{fig:Degree distribution}Quantitative analysis of degree-stratified behavioral segregation. (a) Initial degree distribution follows the expected Poisson form with uniform behavioral composition across degree classes. (b) Steady-state distribution reveals systematic behavioral sorting where high-degree nodes ($k \geq 4$) become predominantly non-drug-using (NP, green), while isolated nodes ($k = 0$) occur almost exclusively among drug users (DP, pink; DA, red). This emergent stratification creates "connectivity-protected" populations with high social capital and behavioral immunity alongside marginalized populations with limited recovery access. Color coding: NP (green), NA (black), DP (pink), DA (red). Parameters: $\delta_s=0.4$, $\delta_w=0.9$, $\alpha=0.4$, $\beta=0.4$, $\gamma=0.6$, $N=10,000$.}
\end{figure}

\subsection{\textit{Network architecture evolution and parameter space analysis}}

\begin{figure}[H]
\centering
\includegraphics[width=1\linewidth]{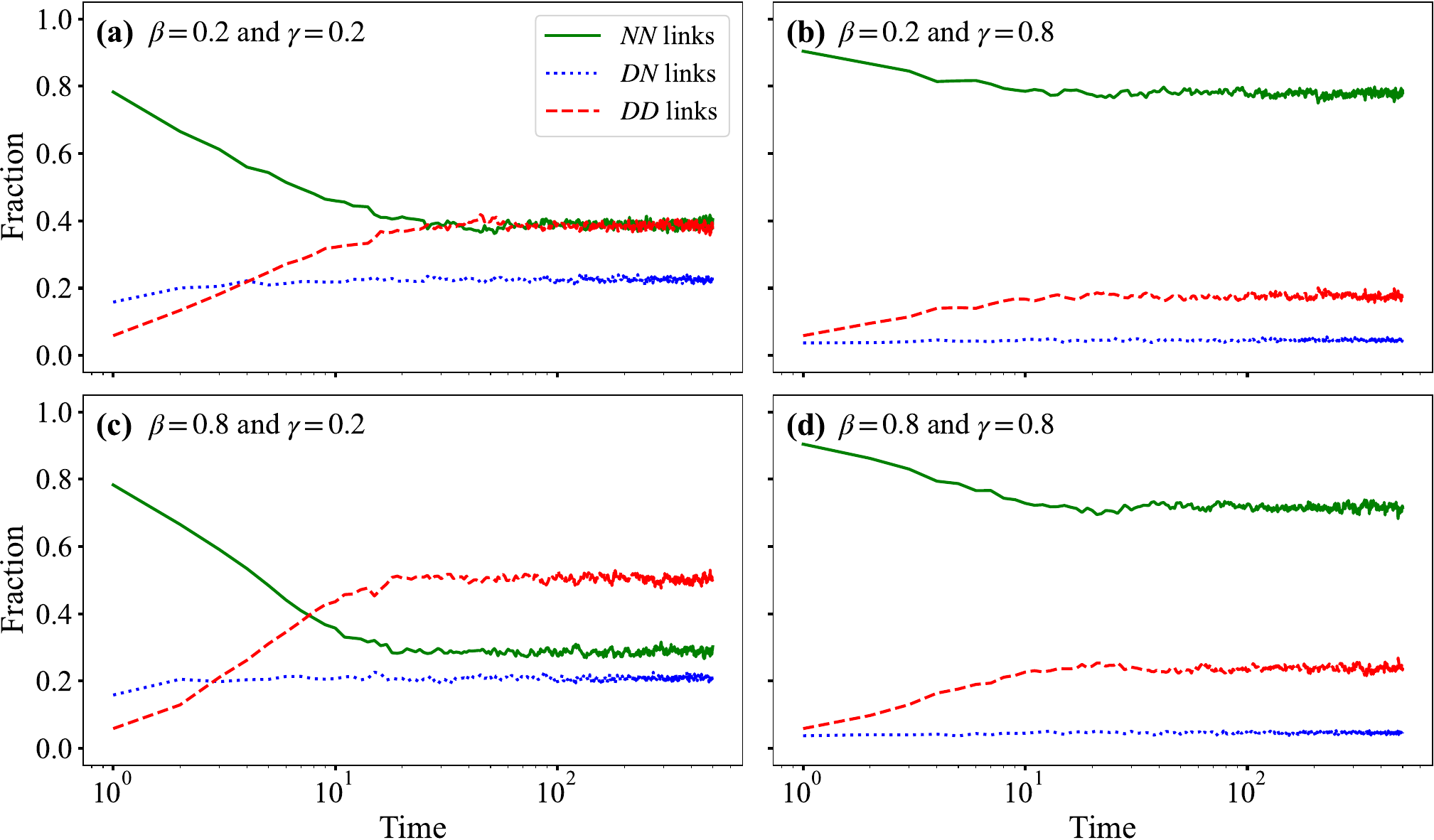}
\caption{\label{fig:links}Dynamic evolution of link composition reveals network adaptation mechanisms. The fraction of link types categorized by behavioral compatibility evolves over time for different addiction conversion rates ($\beta$) and rewiring probabilities ($\gamma$). $NN$ links connect non-drug-using nodes (encompassing NP and NA states), $DD$ links connect drug-using nodes (encompassing DP and DA states), and $DN$ links represent discordant edges subject to rewiring. Parameter combinations: (a) $\beta = 0.2$, $\gamma = 0.2$; (b) $\beta = 0.2$, $\gamma = 0.8$; (c) $\beta = 0.8$, $\gamma = 0.2$; (d) $\beta = 0.8$, $\gamma = 0.8$. The rapid stabilization of $DN$ proportions demonstrates dynamic homeostasis, while $\gamma$ effects dominate over $\beta$, establishing social adaptation as the primary force shaping network architecture.}
\end{figure}

The temporal evolution of link types offers crucial insights into the mechanisms underlying network adaptation, revealing how biological systems can spontaneously organize to limit the spread of pathological behavior. Figure~\ref{fig:links} demonstrates how different parameter combinations shape the relative abundances of homophilic and heterophilic connections over time, revealing the fundamental processes that drive network reorganization. Across all parameter combinations, the proportion of discordant $DN$ links rapidly stabilizes at characteristic equilibrium values, indicating that the system reaches a dynamic balance between link creation and destruction. This equilibrium represents a fundamental property of adaptive biological networks: the continuous turnover of connections maintains stable statistical properties while allowing for individual-level changes, creating a form of "dynamic homeostasis" in the network structure that parallels biological regulatory mechanisms.

Parameter sweeps uncover sharp regime shifts that guide intervention design. When rewiring probability $\gamma$ remains constant, increases in addiction conversion rate $\beta$ systematically shift the link composition toward higher $DD$ proportions at the expense of $NN$ links, reflecting the indirect influence of neurobiological processes on network structure. Conversely, when $\beta$ is held fixed, variations in $\gamma$ produce more dramatic effects on link composition, with higher rewiring rates substantially increasing $NN$ link fractions while reducing $DD$ connections. The magnitude of $\gamma$'s influence significantly exceeds that of $\beta$, establishing social adaptation as the dominant force shaping network structure and highlighting a fundamental principle: behavioral adaptation mechanisms can override biological susceptibility factors in determining population-level outcomes.

\begin{figure}[htbp]
    \centering
    \setlength{\abovecaptionskip}{4pt} 
    \setlength{\belowcaptionskip}{-6pt}
    \begin{subfigure}[t]{0.49\linewidth}
        \centering
        \includegraphics[width=1.04\linewidth]{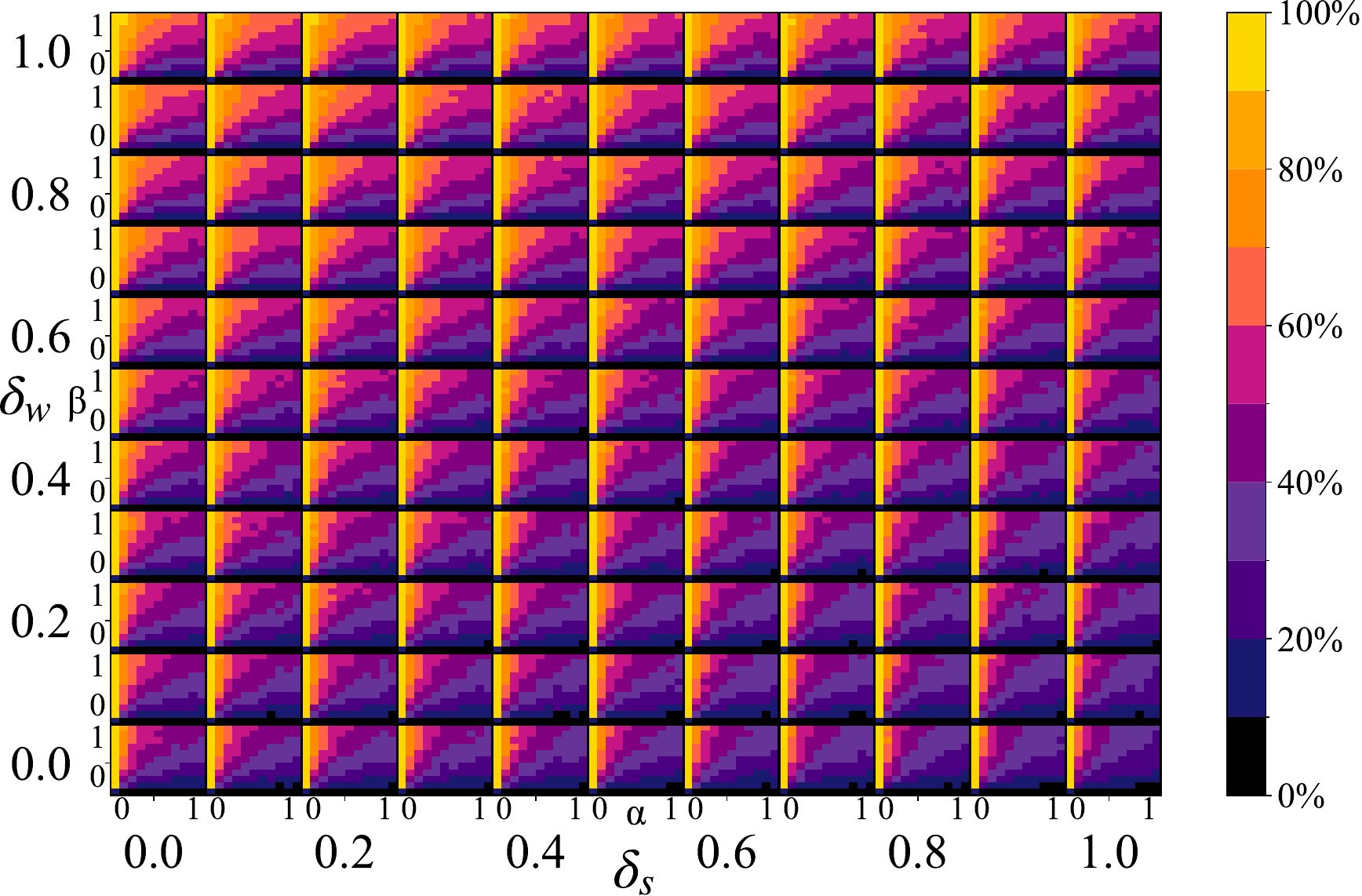}
        \caption{}
    \end{subfigure}
    \hfill
    \begin{subfigure}[t]{0.49\linewidth}
        \centering
        \includegraphics[width=1.05\linewidth]{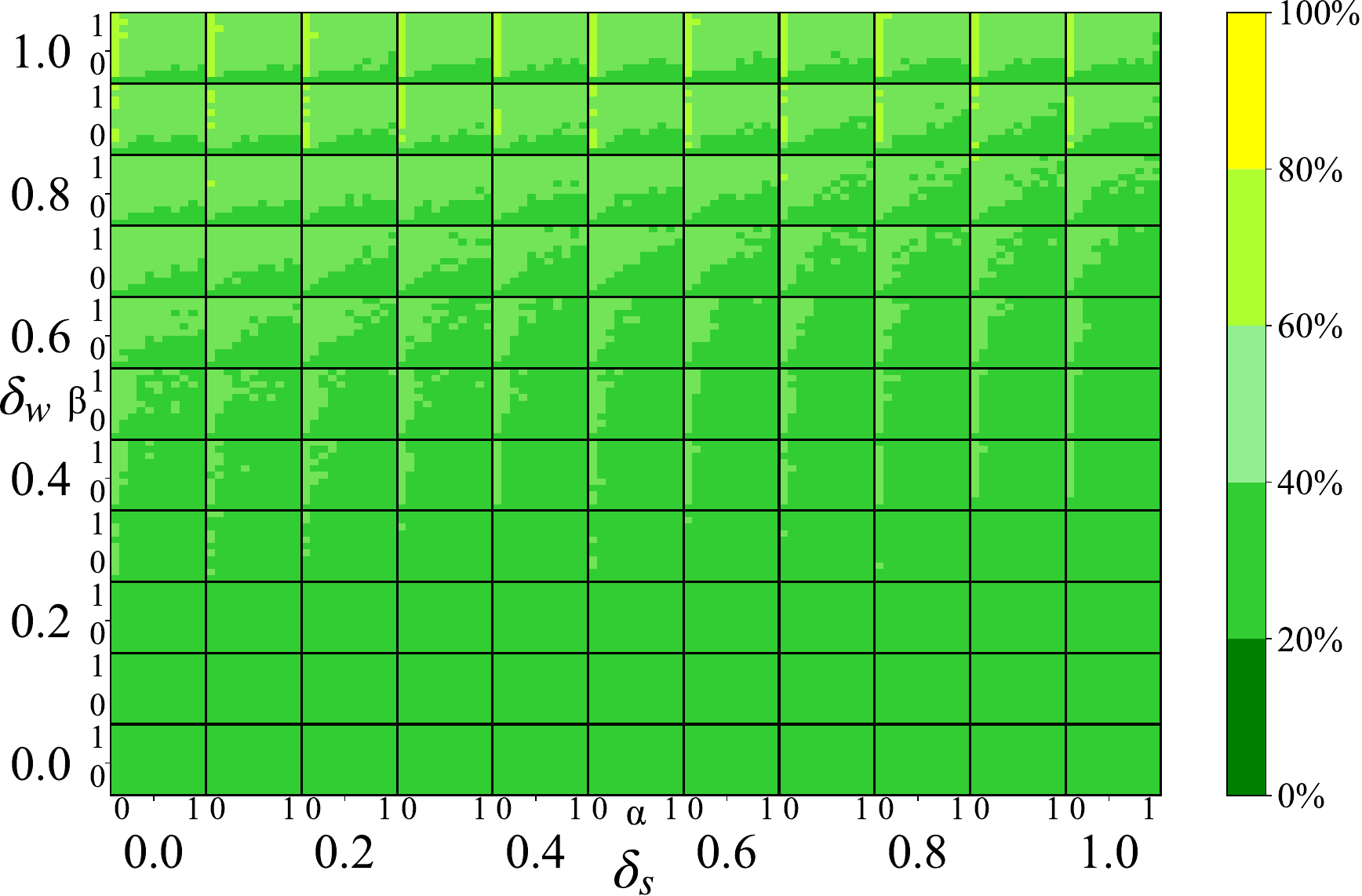}
        \caption{}
    \end{subfigure}
    \caption{\label{fig:11^4}Comprehensive parameter space analysis revealing critical transitions in addiction dynamics. Heat maps show steady-state outcomes across the full $\alpha$-$\beta$ parameter plane, with each mini-panel spanning $0 < \alpha, \beta < 1$. Columns and rows correspond to different values of peer influence ($\delta_s$) and addiction stickiness ($\delta_w$). (a) Addiction fraction shows sharp gradients indicating critical transition boundaries. (b) Drug use fraction displays complex topography with multiple gradient regions. The systematic variation reveals that high recovery rates combined with low addiction conversion consistently produce minimal addiction prevalence, while psychosocial parameters modulate transition sharpness rather than extreme values. Network size $N = 5,000$, $\langle k \rangle = 2$, reference parameters $\delta_s = 0.4$, $\delta_w = 0.9$, $\alpha = 0.4$, $\beta = 0.4$, $\gamma = 0.6$.}
\end{figure}

Our exhaustive exploration of the four-dimensional parameter space, comprising $11^4 = 14,641$ simulations, reveals a complex landscape of behavioral outcomes characterized by sharp phase transitions and nonlinear parameter interactions that mirror critical phenomena in physical systems. Figure~\ref{fig:11^4} presents two-dimensional projections onto the $\alpha$-$\beta$ plane, with systematic variation of peer influence strength $\delta_s$ and addiction stickiness $\delta_w$, providing the most comprehensive characterization of behavioral epidemic dynamics in adaptive biological networks to date. The addiction fraction landscape exhibits pronounced sensitivity to the recovery-addiction axis, with sharp gradients indicating critical transition regions where small parameter changes trigger dramatic shifts in population outcomes, reminiscent of phase transitions in statistical physics.

\subsection{\textit{Individual parameter effects and synergistic intervention strategies}}

\begin{figure}[H]
\centering
\includegraphics[width=1\linewidth]{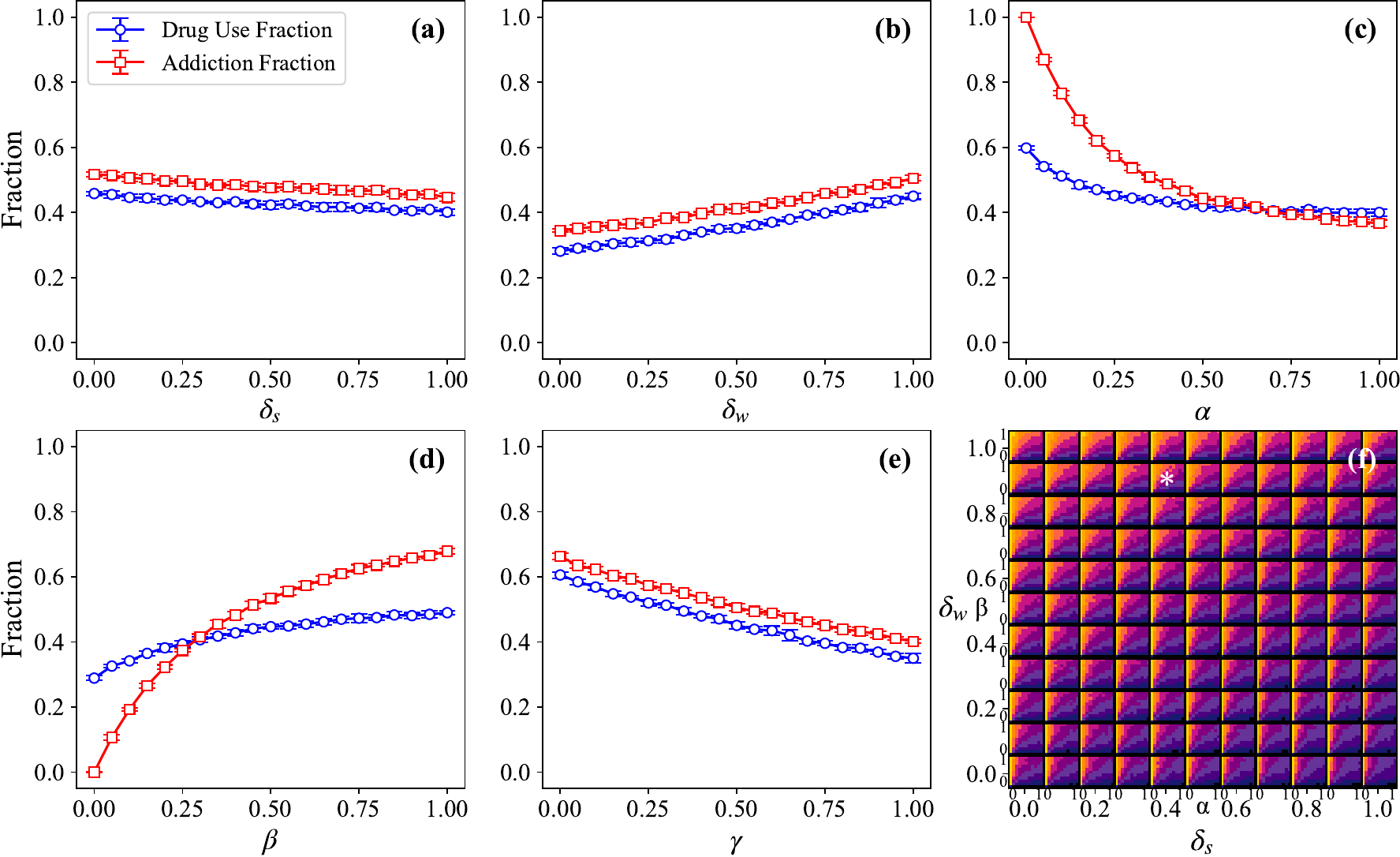}
\caption{\label{fig:one-dimensional}Systematic analysis of individual parameter effects reveals intervention hierarchy. One-dimensional parameter sweeps show steady-state drug use and addiction fractions with other parameters fixed at $\delta_s = 0.4$, $\delta_w = 0.9$, $\alpha = 0.4$, $\beta = 0.4$, $\gamma = 0.6$. Parameters varied: (a) peer influence $\delta_s$, (b) addiction stickiness $\delta_w$, (c) recovery rate $\alpha$, (d) addiction conversion $\beta$, (e) rewiring probability $\gamma$. Panel (f) indicates parameter subset position within broader four-dimensional space, with the star marking the fixed parameter values ($\delta_s = 0.4$, $\delta_w = 0.9$, $\alpha = 0.4$, $\beta = 0.4$, $\gamma = 0.6$) used in panels (a)-(e). Error bars represent standard deviation over 5,000 agents across 30 independent Monte Carlo runs. The rewiring parameter $\gamma$ produces the most dramatic effects, establishing adaptive network management as the most powerful intervention target, while neurobiological parameters $\alpha$ and $\beta$ show characteristic threshold behaviors with maximal sensitivity in lower ranges.}
\end{figure}

To isolate the independent contributions of each biological and social mechanism, we conducted systematic one-dimensional parameter sweeps that establish a clear hierarchy of intervention effectiveness and reveal fundamental principles governing adaptive biological networks. Figure~\ref{fig:one-dimensional} demonstrates the distinct roles of different mechanisms in determining population-level outcomes, with crucial implications for designing effective interventions in biological systems. Peer influence strength $\delta_s$ and addiction stickiness $\delta_w$ exhibit relatively modest effects on steady-state prevalence, with gradual responses suggesting that while these psychosocial factors influence individual decision-making processes, their population-level impacts are constrained by network adaptation processes that create protective clustering effects.

In contrast, the neurobiological parameters $\alpha$ and $\beta$ exhibit pronounced nonlinear effects with characteristic saturation behaviors, revealing fundamental properties of biological systems under intervention. Both parameters exhibit maximal sensitivity in their lower ranges ($0 < \alpha, \beta < 0.5$), followed by diminishing marginal effects at higher values, suggesting that modest improvements in recovery rates or reductions in addiction conversion can yield disproportionate population benefits when starting from high-risk baselines. This threshold behavior indicates that early intervention and prevention programs may be particularly effective in biological systems, representing a general principle for timing interventions in adaptive networks.

Most remarkably, the rewiring parameter $\gamma$ produces the most dramatic effects, with monotonic reductions in both drug use and addiction prevalence as rewiring rates increase across the entire parameter domain. This establishes adaptive network management as the most powerful single intervention target in biological systems, demonstrating that enhanced social adaptation capabilities can enable populations to escape harmful influences while building protective connections. The finding that social adaptation mechanisms can override neurobiological susceptibility factors represents a fundamental insight into the relative importance of behavioral versus biological factors in determining health outcomes at the population level.

\begin{figure}[htbp]
    \centering
    \begin{subfigure}[t]{0.48\linewidth}
        \centering
        \includegraphics[width=\linewidth]{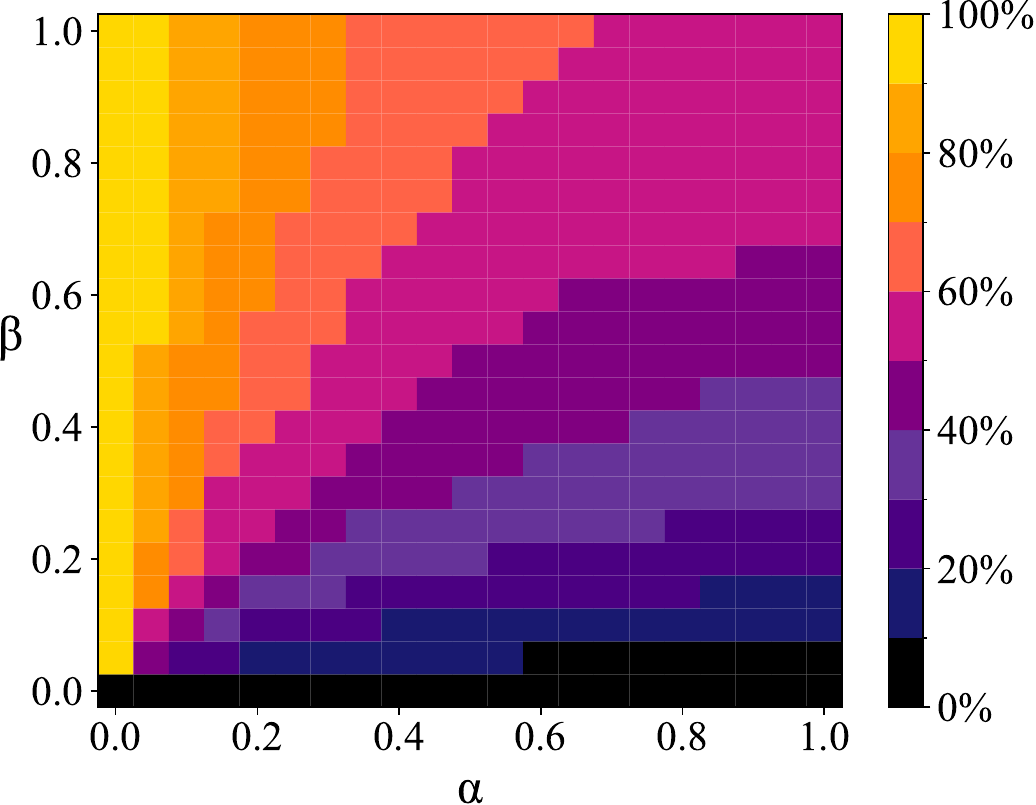}
        \caption{}
    \end{subfigure}
    \hfill
    \begin{subfigure}[t]{0.48\linewidth}
        \centering
        \includegraphics[width=\linewidth]{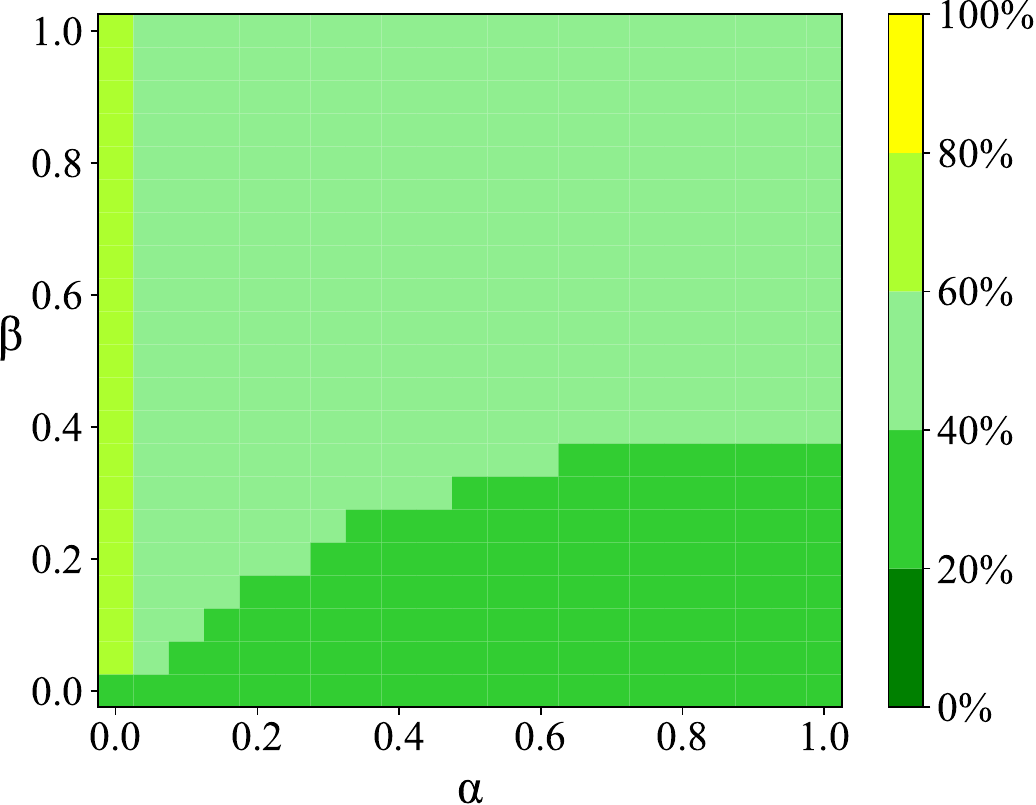}
        \caption{}
    \end{subfigure}
    \caption{\label{fig:two-dimensional}Two-dimensional parameter analysis reveals synergistic effects in neurobiological interventions. Heat maps show steady-state (a) addiction and (b) drug use fractions across the full $(\alpha, \beta)$ parameter space with fixed $\delta_s = 0.4$, $\delta_w = 0.9$, $\gamma = 0.6$. Each grid cell represents averaged outcomes over 5,000 agents across 30 independent simulations. The diagonal symmetry indicates equivalent population outcomes can be achieved through different parameter combinations, providing multiple pathways for intervention design. Sharp diagonal gradients in addiction fraction define critical transition boundaries, while drug use fraction shows more complex topography reflecting broader influences on behavioral choices. The enhanced sensitivity throughout parameter domains demonstrates synergistic interactions between recovery and prevention mechanisms.}
\end{figure}

The identification of $\alpha$ and $\beta$ as highly influential neurobiological parameters motivates a detailed investigation of their joint effects through a two-dimensional analysis, which reveals complex synergistic interactions. Figure~\ref{fig:two-dimensional} presents high-resolution heat maps across the full $\alpha$-$\beta$ parameter space, revealing interaction patterns that inform optimal intervention design strategies for biological systems. The addiction fraction landscape exhibits sharp diagonal gradients that define critical transition boundaries, with enhanced sensitivity throughout the entire parameter domain, reflecting synergistic interactions where the benefits of high recovery rates are amplified in low addiction conversion regimes. This suggests that combined interventions targeting both recovery enhancement and addiction prevention may be more effective than single-component approaches, representing a general principle for intervention design in complex biological systems.

The drug use fraction shows more complex topography with multiple gradient regions and plateau areas, reflecting the broader range of factors that influence behavioral choices beyond simple neurobiological addiction cycles. The diagonal symmetry in both heat maps indicates that equivalent population outcomes can be achieved through different combinations of $\alpha$ and $\beta$ values, suggesting multiple pathways for intervention design that provide flexibility based on feasibility and resource constraints. This finding demonstrates that high recovery rates can partially compensate for high addiction conversion rates and vice versa, offering strategic options for addressing addiction epidemics under varying biological and social constraints. These results collectively establish fundamental principles for understanding how individual-level biological and social mechanisms combine to produce population-level outcomes in adaptive networks, revealing the complex interplay between neurobiological processes and social adaptation that characterizes real biological systems.

\section{\textbf{Discussion}}

What emerges from our comprehensive analysis is a striking revelation about behavioral contagion in adaptive networks. Our results reveal behaviors that classic epidemic models overlook, including feedback characteristic of complex adaptive systems. They seamlessly bridge the fields of statistical physics and biological dynamics. The spontaneous birth of degree-stratified segregation unveils a remarkable form of self-organization. Many local rewiring moves create a clear global hierarchy with no central control. This extraordinary phenomenon illuminates how individual behavioral choices sculpt macroscopic order through collective dynamics. It forges an unexpected parallel to phase separation in physical systems where local molecular interactions generate emergent spatial organization. Statistical mechanics explains the pattern as a balance between social influence and link entropy. The system evolves toward thermodynamically stable configurations, where high-degree nodes accumulate protective behavioral immunity, while low-degree nodes drift toward isolated enclaves of vulnerability. This adaptive network evolution mirrors fundamental principles established in complex systems theory, where self-organization emerges from local interactions without central coordination \cite{yukalov2014self}, paralleling mechanisms observed in diverse biological and social contexts where individual decision-making produces collective patterns \cite{bose2017collective}.

Sharp phase transitions discovered through our exhaustive exploration of $11^4 = 14,641$ parameter combinations reveal behavioral systems operating under principles that mirror critical phenomena in condensed matter physics. These systems exhibit power-law scaling near transition boundaries and extraordinary sensitivity to infinitesimal parameter perturbations. The recovery-addiction axis ($\alpha$-$\beta$) functions as a master order parameter. It wields influence analogous to temperature in thermal phase transitions, where minute threshold crossings trigger catastrophic shifts in population-level outcomes. Small interventions near a boundary can yield large health gains through nonlinear feedback. The degree-dependent stability creates a self-reinforcing mechanism where initial connectivity advantages amplify exponentially over time. This sculpts the observed bifurcated network architecture, which bears a striking resemblance to the phase separation dynamics governing liquid-gas transitions. The critical phenomena we observe align with statistical physics models of behavioral dynamics, where opinion-driven processes can exhibit sharp transitions analogous to physical phase changes \cite{moore2015opinion}. The rewiring rate $\gamma$ is the strongest lever in the model. This reveals a counterintuitive principle in adaptive network theory: topological evolution can completely override intrinsic biological parameters in determining collective fate. Individual capacity for strategic relationship curation proves more decisive than genetic predisposition or pharmacological intervention. The mathematical architecture underlying this dominance operates through exponential scaling of rewiring effects. Cutting a mixed link blocks spread and tightens cohesion among non-users. This generates multiplicative benefits that compound relentlessly over time. Our sophisticated utility-based decision framework marries rational choice theory with stochastic dynamics. Modified Fermi dynamics encode neurobiological asymmetries without sacrificing analytical tractability essential for systematic parameter exploration.

Dynamic homeostasis emerges as a defining characteristic of our adaptive network system. Perpetual microscopic turbulence maintains crystalline macroscopic stability. This creates a biological analog to steady-state thermodynamic systems where continuous energy flows sustain invariant statistical distributions despite underlying molecular chaos. Our system possesses intrinsic self-correcting mechanisms that actively resist perturbations while maintaining organizational coherence. The network exhibits resilience against both stochastic fluctuations and deliberate interventions. This resilience property echoes findings in adaptive coevolutionary networks, where dynamic topology adjustments enhance system stability against perturbations while maintaining functional organization \cite{zhang2020co}. Synergistic coupling between recovery enhancement and addiction prevention demonstrates that carefully orchestrated multi-parameter approaches can achieve transformative outcomes impossible through isolated interventions. Network-based interventions that prioritize enhancing individual social navigation capabilities represent a paradigmatic revolution. This transformative approach acknowledges that empowering individuals to actively architect their social environments can generate protective clustering effects. Mathematical analysis reveals that even modest improvements in rewiring rates can trigger avalanche effects that fundamentally restructure network topology. Yet this adaptive rewiring success story conceals a profound equity paradox. While creating protective sanctuaries for well-connected individuals, the same mechanisms systematically isolate vulnerable populations who lack the necessary social capital. The trade-off reveals that a policy can raise the average yet exacerbate harm for the isolated few. This demands sophisticated bridging strategies that prevent evolutionary drift toward socially inequitable steady states.

The mapped boundaries tell practitioners when and how strongly to intervene. Populations operating near these mathematical knife-edges represent optimal investment opportunities where marginal resource allocation can trigger transformative systemic change. Maximum therapeutic investment should be concentrated at transition boundaries where small perturbations can catalyze large-scale population health improvements. Our framework reveals three synergistic intervention fronts: early prevention through reduced addiction conversion, recovery acceleration through enhanced treatment accessibility, and social navigation training that boosts rewiring capabilities. These intervention strategies reflect broader insights from complex health behavior research, which emphasizes the importance of understanding dynamic interactions between individual cognition and social environment in designing effective behavioral change programs \cite{orr2014complex}. The mathematical framework transcends addiction dynamics to illuminate phenomena where individual adaptation, social influence, and network evolution converge to produce collective outcomes spanning political polarization, technological diffusion, and health behavior adoption.

Degree-based segregation appears whenever links favor similarity, so that the pattern may extend far beyond drugs. This suggests that connectivity-based inequality represents a fundamental organizing principle of adaptive social networks. The critical phenomena observed in our behavioral system establish profound connections to phase transitions in physical systems. Mathematical tools and theoretical frameworks developed in statistical physics may possess unexpected universality across disciplines. Sharp transition boundaries, parameter sensitivity near critical points, and scaling relationships provide compelling evidence that behavioral epidemics obey fundamental laws transcending specific biological or social contexts. This universality extends to political systems, where similar dynamics of behavioral clustering and polarization emerge through individual adaptation mechanisms that reshape social interaction networks \cite{levin2021dynamics}. Traditional approaches that treat network topology as static fundamentally misrepresent the co-evolutionary feedback mechanisms governing long-term outcomes. Our mathematical framework provides sophisticated tools for analyzing these coupled dynamics and identifying intervention leverage points where modest perturbations achieve sustained systemic transformation. The coevolutionary perspective has proven essential in understanding how network structure and dynamics mutually influence each other across diverse biological and social systems \cite{gross2008adaptive}. The scalability of our discoveries across network sizes establishes confidence that these principles govern real-world systems of practical significance. Ultimately, effective addiction policy must work with the grain of natural social rewiring, harnessing collective self-organization to achieve durable population-level resilience through decentralized social adaptation.

The implications of our findings extend far beyond drug use to encompass any behavioral phenomenon where social influence, individual adaptation, and network evolution intersect. This broad applicability reflects the growing recognition that technological innovation and social change often emerge through complex adaptive processes where individual decisions aggregate into collective transformations \cite{frenken2006technological}. Applications span from political polarization and misinformation spread to health behavior adoption and technological innovation diffusion. Our framework reveals three synergistic intervention fronts that leverage natural social dynamics: early prevention through reduced addiction conversion via education and pharmacological prophylaxis, recovery acceleration through enhanced treatment accessibility, and social navigation training that boosts rewiring capabilities by teaching individuals to disengage from harmful influences and cultivate supportive relationships. The model's robust scalability across network sizes and its revelation of distinct mechanistic regimes provide a roadmap for both theoretical advances and empirical validation in real-world social networks.

Looking forward, our work establishes network adaptation as a central explanatory force in behavioral contagion, demanding a paradigm shift where network architecture serves as both diagnostic marker and intervention target. Future extensions incorporating heterogeneous rewiring capabilities, empirical network topologies, and co-diffusing behaviors will deepen our understanding of how economic inequality and social complexity shape addiction ecosystems. The mathematical framework we have developed, bridging rational choice theory with stochastic network dynamics, provides a template for investigating multi-stage behavioral phenomena across diverse domains. Ultimately, our findings demonstrate that effective addiction policy must work with the grain of natural social rewiring rather than against it, harnessing collective self-organization to achieve durable population-level resilience through decentralized social adaptation.

\bibliographystyle{unsrt}
\bibliography{ref}

\begin{thebibliography}{10}

\bibitem{friedman2006addictive}
Herman Friedman, Susan Pross, and Thomas~W Klein.
\newblock Addictive drugs and their relationship with infectious deseases.
\newblock {\em FEMS Immunology \& Medical Microbiology}, 47(3):330--342, 2006.

\bibitem{rosenquist2010spread}
J~Niels Rosenquist, Joanne Murabito, James~H Fowler, and Nicholas~A Christakis.
\newblock The spread of alcohol consumption behavior in a large social network.
\newblock {\em Annals of internal medicine}, 152(7):426--433, 2010.

\bibitem{thombs2019introduction}
Dennis~L Thombs and Cynthia~J Osborn.
\newblock Introduction to addictive behaviors.
\newblock 2019.

\bibitem{khantzian2003understanding}
Edward~J Khantzian.
\newblock Understanding addictive vulnerability: An evolving psychodynamic
  perspective.
\newblock {\em Neuropsychoanalysis}, 5(1):5--21, 2003.

\bibitem{bechara2019neurobehavioral}
Antoine Bechara, Kent~C Berridge, Warren~K Bickel, Jose~A Mor{\'o}n, Sidney~B
  Williams, and Jeffrey~S Stein.
\newblock A neurobehavioral approach to addiction: implications for the opioid
  epidemic and the psychology of addiction.
\newblock {\em Psychological Science in the Public Interest}, 20(2):96--127,
  2019.

\bibitem{hunt2024evolutionary}
Adam Hunt, Giuseppe~Pierpaolo Merola, Tom Carpenter, and Adrian~V Jaeggi.
\newblock Evolutionary perspectives on substance and behavioural addictions:
  Distinct and shared pathways to understanding, prediction and prevention.
\newblock {\em Neuroscience \& Biobehavioral Reviews}, 159:105603, 2024.

\bibitem{masaeli2021prevalence}
Nassim Masaeli and Hadi Farhadi.
\newblock Prevalence of internet-based addictive behaviors during covid-19
  pandemic: A systematic review.
\newblock {\em Journal of addictive diseases}, 39(4):468--488, 2021.

\bibitem{alimoradi2022estimation}
Zainab Alimoradi, Aida Lotfi, Chung-Ying Lin, Mark~D Griffiths, and Amir~H
  Pakpour.
\newblock Estimation of behavioral addiction prevalence during covid-19
  pandemic: a systematic review and meta-analysis.
\newblock {\em Current addiction reports}, 9(4):486--517, 2022.

\bibitem{allen1994some}
Linda~JS Allen.
\newblock Some discrete-time si, sir, and sis epidemic models.
\newblock {\em Mathematical biosciences}, 124(1):83--105, 1994.

\bibitem{tuckwell2007some}
Henry~C Tuckwell and Ruth~J Williams.
\newblock Some properties of a simple stochastic epidemic model of sir type.
\newblock {\em Mathematical biosciences}, 208(1):76--97, 2007.

\bibitem{battista2019modeling}
Nicholas~A Battista, Leigh~B Pearcy, and W~Christopher Strickland.
\newblock Modeling the prescription opioid epidemic.
\newblock {\em Bulletin of mathematical biology}, 81:2258--2289, 2019.

\bibitem{tolles2020modeling}
Juliana Tolles and ThaiBinh Luong.
\newblock Modeling epidemics with compartmental models.
\newblock {\em Jama}, 323(24):2515--2516, 2020.

\bibitem{mayengo2020mathematical}
Maranya~M Mayengo, Gabriel~M Shirima, Snehashish Chakraverty, Moatlhodi
  Kgosimore, Padmanabhan Seshaiyer, and Carmen Caiseda.
\newblock Mathematical modeling of the dynamics of health risks associated with
  alcoholism in tanzania: a literature review.
\newblock {\em Commun. Math. Biol. Neurosci.}, 2020:Article--ID, 2020.

\bibitem{van2022review}
Maarten~WJ Van~den Ende, Sacha Epskamp, Michael~H Lees, Han~LJ van~der Maas,
  Reinout~W Wiers, and Peter~MA Sloot.
\newblock A review of mathematical modeling of addiction regarding both
  (neuro-) psychological processes and the social contagion perspectives.
\newblock {\em Addictive behaviors}, 127:107201, 2022.

\bibitem{huang2014interplay}
Grace~C Huang, Daniel Soto, Kayo Fujimoto, and Thomas~W Valente.
\newblock The interplay of friendship networks and social networking sites:
  Longitudinal analysis of selection and influence effects on adolescent
  smoking and alcohol use.
\newblock {\em American journal of public health}, 104(8):e51--e59, 2014.

\bibitem{karakose2023comprehensive}
Turgut Karakose, Bilal Y{\i}ld{\i}r{\i}m, Tijen T{\"u}l{\"u}ba{\c{s}}, and
  Abdurrahman Kardas.
\newblock A comprehensive review on emerging trends in the dynamic evolution of
  digital addiction and depression.
\newblock {\em Frontiers in psychology}, 14:1126815, 2023.

\bibitem{zhao2024investigating}
Nan Zhao, Guangyu Zhou, Meifen Wei, and David~L Vogel.
\newblock Investigating the cognitive and affective dynamics of social media
  addiction: Insights from peer contexts.
\newblock {\em Journal of Counseling Psychology}, 2024.

\bibitem{nj23101968}
Nicholas John and Nishant Malik.
\newblock Automated discovery of analytical models for epidemic dynamics on
  coevolving networks.
\newblock {\em Journal of Computational Science}, 67:101968, 2023.

\bibitem{marceau2010adaptive}
Vincent Marceau, Pierre-Andr{\'e} No{\"e}l, Laurent H{\'e}bert-Dufresne,
  Antoine Allard, and Louis~J Dub{\'e}.
\newblock Adaptive networks: Coevolution of disease and topology.
\newblock {\em Physical Review E—Statistical, Nonlinear, and Soft Matter
  Physics}, 82(3):036116, 2010.

\bibitem{malik2016transitivity}
Nishant Malik, Feng Shi, Hsuan-Wei Lee, and Peter~J Mucha.
\newblock Transitivity reinforcement in the coevolving voter model.
\newblock {\em Chaos: An Interdisciplinary Journal of Nonlinear Science},
  26(12), 2016.

\bibitem{lee2018evolutionary}
Hsuan-Wei Lee, Nishant Malik, and Peter~J Mucha.
\newblock Evolutionary prisoner’s dilemma games coevolving on adaptive
  networks.
\newblock {\em Journal of complex networks}, 6(1):1--23, 2018.

\bibitem{nmrole2013}
Nishant Malik and Peter~J. Mucha.
\newblock Role of social environment and social clustering in spread of
  opinions in coevolving networks.
\newblock {\em Chaos: An Interdisciplinary Journal of Nonlinear Science},
  23(4):043123, November 2013.

\bibitem{kirmayer2009community}
Laurence~J Kirmayer, Megha Sehdev, Rob Whitley, St{\'e}phane~F Dandeneau, and
  Colette Isaac.
\newblock Community resilience: Models, metaphors and measures.
\newblock {\em International Journal of Indigenous Health}, 5(1):62--117, 2009.

\bibitem{johnson2015infectious}
Pieter~TJ Johnson, Jacobus~C De~Roode, and Andy Fenton.
\newblock Why infectious disease research needs community ecology.
\newblock {\em Science}, 349(6252):1259504, 2015.

\bibitem{bardosh2017addressing}
Kevin~Louis Bardosh, Sadie~J Ryan, Kris Ebi, Susan Welburn, and Burton Singer.
\newblock Addressing vulnerability, building resilience: community-based
  adaptation to vector-borne diseases in the context of global change.
\newblock {\em Infectious diseases of poverty}, 6:1--21, 2017.

\bibitem{choi2019neurobiological}
Jung-Seok Choi, Daniel~Luke King, and Young-Chul Jung.
\newblock Neurobiological perspectives in behavioral addiction, 2019.

\bibitem{valentino2020neurobiology}
Rita~J Valentino, Walter Koroshetz, and Nora~D Volkow.
\newblock Neurobiology of the opioid epidemic: basic and translational
  perspectives.
\newblock {\em Biological psychiatry}, 87(1):2--3, 2020.

\bibitem{volkow2023substance}
Nora~D Volkow and Carlos Blanco.
\newblock Substance use disorders: a comprehensive update of classification,
  epidemiology, neurobiology, clinical aspects, treatment and prevention.
\newblock {\em World Psychiatry}, 22(2):203--229, 2023.

\bibitem{di2012statistical}
Riccardo Di~Clemente and Luciano Pietronero.
\newblock Statistical agent based modelization of the phenomenon of drug abuse.
\newblock {\em Scientific Reports}, 2(1):532, 2012.

\bibitem{d2015statistical}
Maria~R D'Orsogna and Matja{\v{z}} Perc.
\newblock Statistical physics of crime: A review.
\newblock {\em Physics of life reviews}, 12:1--21, 2015.

\bibitem{crokidakis2021modeling}
Nuno Crokidakis and Lucas Sigaud.
\newblock Modeling the evolution of drinking behavior: A statistical physics
  perspective.
\newblock {\em Physica A: Statistical Mechanics and its Applications},
  570:125814, 2021.

\bibitem{bergstrom2024human}
Carl~T Bergstrom and William~P Hanage.
\newblock Human behavior and disease dynamics.
\newblock {\em Proceedings of the National Academy of Sciences},
  121(1):e2317211120, 2024.

\bibitem{grabowski2012relationship}
A~Grabowski and M~Rosi{\'n}ska.
\newblock The relationship between human behavior and the process of epidemic
  spreading in a real social network.
\newblock {\em The European Physical Journal B}, 85:1--6, 2012.

\bibitem{lee2019social}
Hsuan-Wei Lee, Nishant Malik, Feng Shi, and Peter~J Mucha.
\newblock Social clustering in epidemic spread on coevolving networks.
\newblock {\em Physical Review E}, 99(6):062301, 2019.

\bibitem{nunner2022health}
Hendrik Nunner, Vincent Buskens, Alexandra Teslya, and Mirjam Kretzschmar.
\newblock Health behavior homophily can mitigate the spread of infectious
  diseases in small-world networks.
\newblock {\em Social Science \& Medicine}, 312:115350, 2022.

\bibitem{lloyd1996spatial}
Alun~L Lloyd and Robert~M May.
\newblock Spatial heterogeneity in epidemic models.
\newblock {\em Journal of theoretical biology}, 179(1):1--11, 1996.

\bibitem{real2007spatial}
Leslie~A Real and Roman Biek.
\newblock Spatial dynamics and genetics of infectious diseases on heterogeneous
  landscapes.
\newblock {\em Journal of the Royal Society Interface}, 4(16):935--948, 2007.

\bibitem{xiong2024interacting}
Yunfeng Xiong, Chuntian Wang, and Yuan Zhang.
\newblock Interacting particle models on the impact of spatially heterogeneous
  human behavioral factors on dynamics of infectious diseases.
\newblock {\em PLOS Computational Biology}, 20(8):e1012345, 2024.

\bibitem{si2025evolution}
Zehua Si, Takayuki Ito, and Hsuan-Wei Lee.
\newblock Evolution of cooperation in spatial public goods games with migration
  and interactive diversity.
\newblock {\em Applied Mathematics and Computation}, 506:129544, 2025.

\bibitem{henneberger2021peer}
Angela~K Henneberger, Dawnsha~R Mushonga, and Alison~M Preston.
\newblock Peer influence and adolescent substance use: A systematic review of
  dynamic social network research.
\newblock {\em Adolescent research review}, 6(1):57--73, 2021.

\bibitem{duncan2019fast}
Jacob~P Duncan, Teresa Aubele-Futch, and Monica McGrath.
\newblock A fast-slow dynamical system model of addiction: Predicting relapse
  frequency.
\newblock {\em SIAM Journal on Applied Dynamical Systems}, 18(2):881--903,
  2019.

\bibitem{field2020recovery}
Matt Field, Nick Heather, James~G Murphy, Tom Stafford, Jalie~A Tucker, and
  Katie Witkiewitz.
\newblock Recovery from addiction: Behavioral economics and value-based
  decision making.
\newblock {\em Psychology of Addictive Behaviors}, 34(1):182, 2020.

\bibitem{matthews2017stigma}
Steve Matthews, Robyn Dwyer, and Anke Snoek.
\newblock Stigma and self-stigma in addiction.
\newblock {\em Journal of bioethical inquiry}, 14:275--286, 2017.

\bibitem{traulsen2009stochastic}
Arne Traulsen and Christoph Hauert.
\newblock Stochastic evolutionary game dynamics.
\newblock {\em Reviews of nonlinear dynamics and complexity}, 2:25--61, 2009.

\bibitem{altrock2009deterministic}
Philipp~M Altrock and Arne Traulsen.
\newblock Deterministic evolutionary game dynamics in finite populations.
\newblock {\em Physical Review E—Statistical, Nonlinear, and Soft Matter
  Physics}, 80(1):011909, 2009.

\bibitem{robert2016processes}
Marion Robert, Alban Thomas, and Jacques-Eric Bergez.
\newblock Processes of adaptation in farm decision-making models. a review.
\newblock {\em Agronomy for sustainable development}, 36:1--15, 2016.

\bibitem{lee2025granular}
Hsuan-Wei Lee and Yi-Ning Weng.
\newblock Granular q-learning adaptation boosts collective welfare in
  multi-agent prisoner’s dilemma.
\newblock {\em Chaos, Solitons \& Fractals}, 199:116642, 2025.

\bibitem{tortia2025stakeholders}
Ermanno~C Tortia.
\newblock Stakeholders self-organization and adaptive governance in social
  enterprises: Membership representation, worker control and client
  orientation.
\newblock {\em Systems Research and Behavioral Science}, 42(2):396--410, 2025.

\bibitem{lee2025enhancing}
Hsuan-Wei Lee, Szu-Ping Chen, and Feng Shi.
\newblock Enhancing cooperation in dynamic networks through
  reinforcement-learning-based rewiring strategies.
\newblock {\em New Journal of Physics}, 2025.

\bibitem{baldassarri2007dynamics}
Delia Baldassarri and Peter Bearman.
\newblock Dynamics of political polarization.
\newblock {\em American sociological review}, 72(5):784--811, 2007.

\bibitem{liu2018investigating}
Shiyong Liu, Hong Xue, Yan Li, Judy Xu, and Youfa Wang.
\newblock Investigating the diffusion of agent-based modelling and system
  dynamics modelling in population health and healthcare research.
\newblock {\em Systems Research and Behavioral Science}, 35(2):203--215, 2018.

\bibitem{ahmad2023households}
Munir Ahmad, Irfan Khan, Muhammad Qaiser~Shahzad Khan, Gul Jabeen, Hafiza~Samra
  Jabeen, and Cem I{\c{s}}{\i}k.
\newblock Households' perception-based factors influencing biogas adoption:
  Innovation diffusion framework.
\newblock {\em Energy}, 263:126155, 2023.

\bibitem{lu2024agents}
Hsiu-Chi Lu and Hsuan-wei Lee.
\newblock Agents of discord: Modeling the impact of political bots on opinion
  polarization in social networks.
\newblock {\em Social Science Computer Review}, page 08944393241270382, 2024.

\bibitem{szabo2005phase}
Gy{\"o}rgy Szab{\'o}, Jeromos Vukov, and Attila Szolnoki.
\newblock Phase diagrams for an evolutionary prisoner’s dilemma game on
  two-dimensional lattices.
\newblock {\em Physical Review E—Statistical, Nonlinear, and Soft Matter
  Physics}, 72(4):047107, 2005.

\bibitem{perc2010coevolutionary}
Matja{\v{z}} Perc and Attila Szolnoki.
\newblock Coevolutionary games—a mini review.
\newblock {\em BioSystems}, 99(2):109--125, 2010.

\bibitem{yukalov2014self}
Vyacheslav~I Yukalov and Didier Sornette.
\newblock Self-organization in complex systems as decision making.
\newblock {\em Advances in Complex Systems}, 17(03n04):1450016, 2014.

\bibitem{bose2017collective}
Thomas Bose, Andreagiovanni Reina, and James~AR Marshall.
\newblock Collective decision-making.
\newblock {\em Current opinion in behavioral sciences}, 16:30--34, 2017.

\bibitem{moore2015opinion}
Thomas~W Moore, Patrick~D Finley, Benjamin~J Apelberg, Bridget~K Ambrose,
  Nancy~S Brodsky, Theresa~J Brown, Corinne Husten, and Robert~J Glass.
\newblock An opinion-driven behavioral dynamics model for addictive behaviors.
\newblock {\em The European Physical Journal B}, 88:1--28, 2015.

\bibitem{zhang2020co}
Huixin Zhang, Xueming Liu, Qi~Wang, Weidong Zhang, and Jianxi Gao.
\newblock Co-adaptation enhances the resilience of mutualistic networks.
\newblock {\em Journal of the Royal Society Interface}, 17(168):20200236, 2020.

\bibitem{orr2014complex}
Mark~G Orr and David~C Plaut.
\newblock Complex systems and health behavior change: insights from cognitive
  science.
\newblock {\em American journal of health behavior}, 38(3):404--413, 2014.

\bibitem{levin2021dynamics}
Simon~A Levin, Helen~V Milner, and Charles Perrings.
\newblock The dynamics of political polarization, 2021.

\bibitem{gross2008adaptive}
Thilo Gross and Bernd Blasius.
\newblock Adaptive coevolutionary networks: a review.
\newblock {\em Journal of the Royal Society Interface}, 5(20):259--271, 2008.

\bibitem{frenken2006technological}
Koen Frenken.
\newblock Technological innovation and complexity theory.
\newblock {\em Economics of Innovation and New Technology}, 15(2):137--155,
  2006.

\end{thebibliography}

\end{document}